\pdfoutput=1

\documentclass[conference]{IEEEtran}

\IEEEoverridecommandlockouts 

\usepackage[cmex10]{amsmath}
\usepackage{graphicx}
\usepackage{cite}
\usepackage{amsfonts,amssymb,bbm}
\usepackage{mathtools} 
\usepackage{color}
\usepackage{enumerate}

\def\calX{{\mathcal{X}}}

\DeclareMathAlphabet{\CMmathcal}{OMS}{cmsy}{m}{n}
\renewcommand{\mathcal}[1]{\CMmathcal{#1}}

\newcommand{\eqtop}[1]{\overset{(#1)}{=}}
\newcommand{\leqtop}[1]{\overset{(#1)}{\leq}}
\newcommand{\geqtop}[1]{\overset{(#1)}{\geq}}

\newtheorem{theorem}{Theorem}
\newtheorem{lemma}{Lemma}

\newtheorem{definition}{Definition}
\newtheorem{remark}{Remark}

\newcommand*{\QEDD}{\hfill\ensuremath{\Diamond}}%
\newcommand*{\QEDA}{\hfill\IEEEQED}%
\newcommand*{\QEDB}{\hfill\IEEEQEDopen}%

\begin{document}

\title{Authentication With a Guessing Adversary}

\author{\IEEEauthorblockN{Farshad~Naghibi, Tobias~J.~Oechtering, and Mikael~Skoglund}
    \IEEEauthorblockA{Department of Communication Theory \\
    School of Electrical Engineering and ACCESS Linnaeus Center \\
    KTH Royal Institute of Technology, SE-100 44 %
    Stockholm, Sweden}
    Emails: \{farshadn,oech,skoglund\}@kth.se
    }

\maketitle

\begin{abstract}
In this paper, we consider the authentication problem where a candidate measurement presented by an unidentified user is compared to a previously stored measurement of the legitimate user, the \emph{enrollment}, with respect to a certain distortion criteria for authentication. An adversary wishes to impersonate the legitimate user by guessing the enrollment until the system authenticates him. For this setting, we study the minimum number of required guesses (on average) by the adversary for a successful impersonation attack and find the complete characterization of the asymptotic exponent of this metric, referred to as the \emph{deception exponent}. Our result 
is a direct application of the results of the
Guessing problem by Arikan and Merhav [19]. 
Paralleling the work in [19] 
we also extend this
result to the case where the adversary may have access to additional
side information correlated to the enrollment data.

The paper is a revised version of a submission to \emph{IEEE WIFS
  2015}, with the referencing to the paper [19] 
clarified compared with the conference version.
\end{abstract}

\section{Introduction}

Societal development will lead to changes in the way communication systems are used. Devices are becoming more and more connected to one another through massive distributed networks. These devices may contain sensitive data such as personal, commercial, or even military information. Therefore, securing access to such contents using authentication techniques that limits the access to legitimate users is of high importance, especially with the increasing rate at which manipulating digital information with complicated methods, which require only low-cost systems, becomes easier. 

\begin{figure}[t]
 \centering
 \includegraphics[width=\columnwidth]{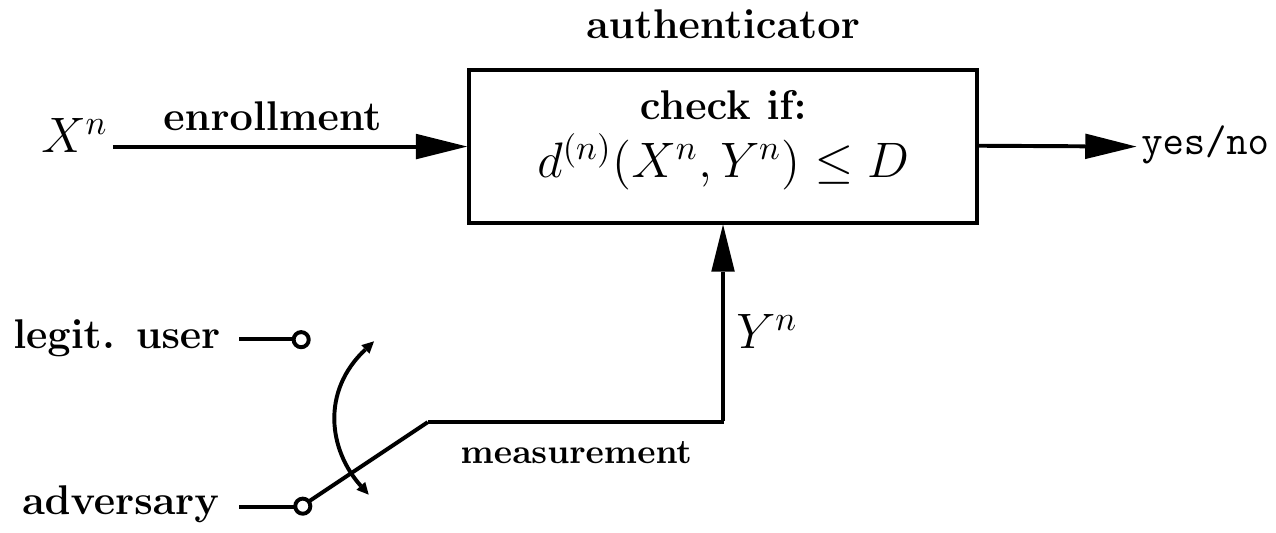}
 \caption{Illustration of the authentication system.}
 \label{fig:sysModel}
\end{figure}

Motivated by these concerns, we study an authentication problem, depicted in Figure~\ref{fig:sysModel}, from an information-theoretic perspective. In the authentication system considered in this paper, a candidate measurement presented by an unidentified user for authentication is compared to a previously stored measurement of the legitimate user which is referred to as the \emph{enrollment}. The output of the system is a binary decision that identifies whether the user is legitimate or malicious. This authentication system can be used in variety of applications such as biometric systems (see e.g., [1] and [2]), 
where measurements are biometric features like fingerprints, or, more generally, it can be used in high-dimensional database systems (see e.g., [3]). 

Due to the inherent randomness in the measurement environment or the use of different hardware, measurements are often noisy. Therefore, the authentication system should be designed such that it tolerates a certain level of distortion between the measurement in the authentication phase and the enrollment phase.

We assume that there exists a malicious user whose goal is to deceive the authentication system by impersonating the legitimate user. This malicious user can potentially have access to some additional side information correlated to the enrollment data. The strategy of the adversary is to present a series of \emph{guesses} of the enrollment until he is authenticated, i.e., the distortion between his guess and the enrollment falls below the required distortion limit which is predetermined in the system. In this paper, we provide an assessment of the vulnerability of such authentication system.

\subsection{Related Work and Contributions}

In authentication problems, one of the main information-theoretic performance metrics of interest is the success probability of the adversary, i.e., the probability of false-acceptance in the system. The initial work on characterizing the probability of false-acceptance is [4], 
where the authors studied a system in which the enrollment data is compressed and stored to be compared with later measurements by the authenticator with respect to a certain distortion criteria. In the case of independent measurement and enrollment data, the trade-off between the compression rate and the exponent of the false-acceptance probability was characterized. 
However, for the case of correlated measurement and enrollment data, which corresponds to availability of additional side information at the adversary, inner and outer bounds were derived which did not match. %

Another approach for authentication was considered in [5] and [6],
where the authentication problem was modeled as a statistical hypothesis testing problem. Then, the exponent of false-acceptance probability was bounded using a large deviation approach.

A closely related problem to [4] 
has been investigated in the literature in the context of database queries, where the focus is not on authenticating a user, but to identify if a given measurement is similar to the compressed enrollment stored in the database. Authors in [7]
characterized the maximum rate of enrollments that can be reliably identified in the database, and extended the results in [8]
to the case where reconstruction of the enrollment is also required. In addition, for some special cases of this problem, the exponent of the false-acceptance probability is characterized. These results can be found in [3]
for the quadratic Gaussian case, in [9] 
for the case when no distortion is allowed, i.e., exact match is required, and in [10]
for the case of binary symmetric source with Hamming distortion.

The authors in [11] 
studied the authentication problem in the context of biometric systems and determined the exponent of the false-acceptance probability with compressed and uncompressed enrollment data when the adversary has no side information. For the case of an adversary with side information and uncompressed enrollment, the authors in [12]
derived the optimal false-acceptance exponent.

In a different line of work, authentication is considered in a communication setting, when a user transmits a message to a receiver. This type of authentication is referred to as \emph{message authentication}. Here authentication is performed using a shared secret key which is not available to the adversary. The study of information-theoretic performance limits for this setting was initiated in [13] 
and has been followed up to this date, see for instance [14-18] 
and references therein. However, we do not consider message authentication or the use of a secret key in our current study.

In this paper, we study the authentication problem with respect to a
certain distortion limit in which the adversary tries to authenticate
himself in the system by guessing the (uncompressed) enrollment
sequence. Our performance metric is different from those considered in
the prior works in authentication; We consider the average number of
guesses that the adversary needs to make to successfully deceive the
authentication system, and completely determine the asymptotic
exponent of this metric. 
Our work is a direct application of
the results of the Guessing problem in [19]. 
This result is of interest in the design of the authentication systems as
it characterizes the minimum required complexity (on average) of an
adversary for successful deception in the system. Furthermore, we
extend our result to the case where the adversary has access to
additional side information correlated to the enrollment.

\subsection{Notations and Organization}

In this paper, we use capital letters to indicate a random variable, small letters to indicate realization of a random variable, calligraphic letters to denote a set, e.g., $\calX$, and $|\calX|$ to indicate the cardinality of the set. The notation $X^n$ denotes the length-$n$ sequence $\{X_1,\dots,X_n\}$. A probability distribution on $\calX$ is denoted by $\mathsf{P}$, i.e., $\mathsf{P}\coloneqq \{\mathsf{P}(x),x\in\calX\}$. Moreover, the probability of the sequence $x^n$ is denoted by $\mathsf{P}^n(x^n)$.

The rest of the paper is organized as follows: In Section~\ref{sec:probForm}, we describe the problem settings along with the definitions. The main result on the characterization of the deception exponent is presented in Section~\ref{sec:main_result} and the proofs are given in Section~\ref{sec:proof_th1}. In Section~\ref{sec:auth_SI}, we consider an extension of the problem in which the adversary has access to some additional side information correlated to the enrollment. Finally, Section~\ref{sec:conclusions} concludes the paper.

\vspace{6pt}

\section{Problem Settings and Definitions}
\label{sec:probForm}

Consider an independent and identically distributed (i.i.d.) random memoryless source sequence $X^n$ generated according to the distribution $\mathsf{P}$ defined on a finite alphabet $\calX$. Here, $n$ corresponds to the measurement length and the sequence represents the enrollment in the system based on which users are authenticated. Denote the reconstruction alphabet by $\mathcal{Y}$ and let $d:\calX\times\mathcal{Y}\rightarrow[0,d_{\max}]$ be a finite distortion measure. We define the component-wise mean distortion between two sequences $x^n\in\calX^n$ and $y^n\in\mathcal{Y}^n$ as 
\begin{equation}
d^{(n)}(x^n,y^n)\coloneqq \frac{1}{n}\sum_{i=1}^n d(x_i,y_i).
\end{equation}

Recall that in lossy source coding of a discrete memoryless source $X$ with distribution $\mathsf{Q}$ and a distortion measure $d$, the rate-distortion function $R(D,\mathsf{Q})$ defined as the infimum of all achievable rates for a given distortion $D$ [20, Theorem 7.3]
is given by
\begin{equation}
\label{eq:R_DQ}
R(D,\mathsf{Q}) = \min_{\mathsf{P}(y|x):\mathbb{E}[d(X,Y)]\leq D} I(X;Y),
\end{equation}
where it is assumed that for every $x\in\mathcal{X}$,  there exists a reconstruction symbol $y\in\mathcal{Y}$ such that
\begin{equation}
\label{eq:dmin0}
d(x,y)=0.
\end{equation}

In the authentication system considered in this paper, a user is authenticated as legitimate if the distortion between the provided sequence at the authentication step and the enrollment is below a predefined limit $D$.

Now, assume that a malicious user tries to launch an impersonation attack by guessing the enrollment sequence $x^n$ with a certain strategy as follows. For every measurement length $n$, the adversary has a strategy $\mathcal{S}^{(n)}\subseteq\mathcal{Y}^n$ which is an ordered set of $n$-length sequences (\emph{guesses}) $y^n(j)\in\mathcal{Y}^n$ with $j=\{1,2,\dots\}$, that is, 
\begin{equation}
\mathcal{S}^{(n)}\coloneqq\big\{y^n(1),y^n(2),\dots\big\}.
\end{equation}
Using this strategy, the adversary produces a sequence of guesses until a guess $y^n(j)$ is found such that 
\begin{equation}
d^{(n)}\big(x^n,y^n(j)\big)\leq D,
\end{equation}
which leads to the adversary being authenticated by the system. %
\begin{remark}
There always exists a strategy for the adversary that leads to authentication. This stems from the fact that one strategy could be the set of all possible sequences in the reconstruction domain $\mathcal{Y}^n$, and based on the assumption in \eqref{eq:dmin0}, there is always a guess that satisfies any required distortion constraint $D\geq 0$ for authentication.
\end{remark}

Define the counting function $G^{(n)}(\cdot)$ of the adversary as the function that for a given enrollment sequence $x^n\in\calX^n$ returns the index $j$ of the first guess $y^n(j)\in\mathcal{S}^{(n)}$ such that $d^{(n)}(x^n,y^n(j))\leq D$. If no such a guess exists, it returns $G^{(n)}(x^n)\!=\!\infty$.

\begin{definition}
The \emph{deception exponent} under the distortion constraint $D$ is defined as
\begin{equation}
\label{eq:ED_def}
E(D) \coloneqq \liminf_{n\to\infty}\frac{1}{n}\min_{\mathcal{S}^{(n)}}\log\mathbb{E}_{\mathsf{P}}\big[G^{(n)}(X^n)\big].
\end{equation}
\QEDD
\end{definition}

The above definition characterizes the asymptotic exponent of the average number of guesses that the adversary needs to make with its best possible (optimal) strategy such that he is authenticated by the system. This can be used for adjusting the distortion constraint $D$ so as to limit the adversarial success. 

\vspace{6pt}

\section{Main Result} 
\label{sec:main_result}

The main result of this paper is a complete characterization of the
deception exponent. Since our definition of the deception exponent is
equivalent to a special case of the guessing exponent introduced in
[19], 
the result also follows from [19]. 

\begin{theorem}
\label{th:main_res}
The deception exponent under the distortion constraint $D$ is
\begin{equation}
E(D) = \max_{\mathsf{Q}}\big[R(D,\mathsf{Q})-D(\mathsf{Q}\|\mathsf{P})\big],
\end{equation}
where $R(D,\mathsf{Q})$ is the rate-distortion function defined in \eqref{eq:R_DQ} and $D(\mathsf{Q}\|\mathsf{P})$ is the relative entropy between the type distribution $\mathsf{Q}$ and the source distribution $\mathsf{P}$.
\QEDB
\end{theorem}

The proof of Theorem~\ref{th:main_res}, stated in the next section,
is a special case of
a corresponding proof in [19] 
applied to the
authentication problem formulated here, and considering the first
moment only. 
Even though it follows directly from [19], 
we include the proof for completeness. 
Achievability is proved by constructing adversarial strategies based
on different types on $\mathcal{X}$ and relating the corresponding
type classes to the rate-distortion function using the type covering
lemma [20]. 
The idea in the converse
is a proof by contradiction. That is, for a given distortion
constraint, we construct a rate-distortion code using adversary's
strategy list followed by an entropy encoder and assert that its rate
cannot be smaller than the rate-distortion function, otherwise it
would contradict the converse of the rate-distortion theory.


\vspace{6pt}

\section{Proof of Theorem~\ref{th:main_res}  [19]} 
\label{sec:proof_th1}

\subsection{Achievability}
\label{sec:achievability}

In this section, it is shown that for any distortion constraint $D\geq 0$, there exists a sequence of adversarial strategies $\{\mathcal{S}^{(n)}\}$ such that
\begin{equation}
\label{eq:ach}
\limsup_{n\to\infty}\frac{1}{n}\log\mathbb{E}_{\mathsf{P}}\big[G^{(n)}(X^n)\big] \leq \max_{\mathsf{Q}}\big[R(D,\mathsf{Q})-D(\mathsf{Q}\|\mathsf{P})\big],
\end{equation}
for every distribution $\mathsf{P}$ on $\mathcal{X}$. This inequality
gives an upper bound on $E(D)$.

The proof uses the following lemma referred to as the type covering lemma in [20, Lemma 9.1] 
\begin{lemma}
Let $\mathcal{Q}^{(n)}(\mathcal{X})$ denote the set of all types on $\mathcal{X}$ 
of $n$-length sequences in $\mathcal{X}^n$. Then, for any distortion measure on $\mathcal{X}\times\mathcal{Y}$, any type $\mathsf{Q}\in\mathcal{Q}^{(n)}(\mathcal{X})$, and any distortion level $D\geq 0$, there exists a set $\mathcal{B}\subseteq\mathcal{Y}^n$ such that 
\begin{equation}
d^{(n)}(x^n,\mathcal{B})\coloneqq\min_{y^n\in\mathcal{B}} d^{(n)}(x^n,y^n)\leq D,
\end{equation}
for every $x^n\in T_\mathsf{Q}^n$, and
\begin{equation}
\label{eq:typ_covering}
\frac{1}{n}\log |\mathcal{B}|\leq R(D,\mathsf{Q}) + \delta,
\end{equation}
where $T_\mathsf{Q}^n$ denotes the type class of $\mathsf{Q}$ (set of sequences of type $\mathsf{Q}$ in $\mathcal{X}^n$) and $\delta\to 0$ as $n\to\infty$.
\QEDB
\end{lemma}

This lemma allows us to divide the set of all types of $n$-length sequences, which is the set of all possible guesses, to smaller subsets for each type that can be described by the rate-distortion function. %
Now, we proceed with the proof. For each type $\mathsf{Q}_k\in \mathcal{Q}^{(n)}(\mathcal{X})$ with $k\in\{1,\dots,|\mathcal{Q}^{(n)}(\mathcal{X})|\}$, let $\mathcal{B}_k$ be the corresponding set that satisfies the type covering lemma above. We can now order the types in $\mathcal{Q}^{(n)}(\mathcal{X})$ with respect to their values of rate-distortion function in an increasing order, i.e., for the ordered list $\{\mathsf{Q}_1,\mathsf{Q}_2,\dots\}$, we have $R(D,\mathsf{Q}_k)\leq R(D,\mathsf{Q}_{k+1})$.

The adversary's strategy $\mathcal{S}^{(n)}$ can be constructed as a list which consists of ordered concatenation of the sets $\{\mathcal{B}_1,\mathcal{B}_2,\dots\}$, corresponding to $\{\mathsf{Q}_1,\mathsf{Q}_2,\dots\}$. We have
\begin{align}
\mathbb{E}_{\mathsf{P}}\big[&G^{(n)}(X^n)\big] = \sum_{x^n\in \mathcal{X}^n} \mathsf{P}^n(x^n) G^{(n)}(x^n) \notag\\
&= \sum_{k=1}^{|\mathcal{Q}^{(n)}(\mathcal{X})|} \sum_{x^n\in T_{\mathsf{Q}_k}^n} \mathsf{P}^n(x^n) G^{(n)}(x^n) \notag\\
&\leqtop{a} \sum_{k=1}^{|\mathcal{Q}^{(n)}(\mathcal{X})|} \sum_{x^n\in T_{\mathsf{Q}_k}^n} \mathsf{P}^n(x^n) \sum_{k'\leq k} |\mathcal{B}_{k'}| \notag\\
&\leqtop{b} \sum_{k=1}^{|\mathcal{Q}^{(n)}(\mathcal{X})|} \exp\Big[-nD(\mathsf{Q}_k\|\mathsf{P})\Big] \sum_{k'\leq k} |\mathcal{B}_{k'}| \notag\\
&\leqtop{c} \sum_{\mathsf{Q}\in \mathcal{Q}^{(n)}(\mathcal{X})} \exp\Big[-nD(\mathsf{Q}\|\mathsf{P})\Big] \notag\\ 
&\qquad\qquad\qquad \cdot \exp\Big[n(R(D,\mathsf{Q}) + \delta)\Big] \notag\\
&= \sum_{\mathsf{Q}\in \mathcal{Q}^{(n)}(\mathcal{X})} \exp\Big[n\big(R(D,\mathsf{Q}) - D(\mathsf{Q}\|\mathsf{P}) + \delta\big)\Big] \notag\\
&\leqtop{d} (n+1)^{|\mathcal{X}|} 
\exp\Big[n\big(R(D,\mathsf{Q}) - D(\mathsf{Q}\|\mathsf{P}) + \delta\big)\Big], \label{eq:ach_EG}
\end{align}
where
\begin{enumerate}[(a)]
	\item follows from the definition of the counting function, i.e., its return value is at most the size of the strategy list;
	\item follows from [20, Lemma 2.6], 
	i.e., for every type $\mathsf{Q}$ of sequences in $\mathcal{X}^n$ and distribution $\mathsf{P}$ on $\mathcal{X}$, we have
	\begin{equation}
	\mathsf{P}^n(T_\mathsf{Q})\leq \exp\big[-nD(\mathsf{Q}\|\mathsf{P})\big];
	\end{equation}
	\item follows from the type covering lemma in \eqref{eq:typ_covering};
	\item follows from the type counting lemma [20, Lemma 2.2]. 
\end{enumerate}
Taking the logarithms of both sides of \eqref{eq:ach_EG}, we obtain
\begin{align}
\frac{1}{n}\log\mathbb{E}_{\mathsf{P}}\big[G^{(n)}(X^n)\big] &\leq \frac{|\mathcal{X}|\log(n+1)}{n} \notag\\
&\quad\quad + R(D,\mathsf{Q})-D(\mathsf{Q}\|\mathsf{P}) + \delta.
\end{align}
Since the left-hand side is not dependent on the distribution $\mathsf{Q}$, taking the limit as $n\to\infty$ on both sides and the maximum on the right-hand side over all possible distributions $\mathsf{Q}$, we get \eqref{eq:ach} which concludes the achievability proof.\QEDA

\vspace{6pt}

\subsection{Converse}
\label{sec:converse}

In this section, we prove that for any arbitrary sequence of adversarial strategies $\{\mathcal{S}^{(n)}\}$ with distortion constraint $D\geq 0$, we have
\begin{equation}
\label{eq:converse}
\liminf_{n\to\infty}\frac{1}{n}\log\mathbb{E}_{\mathsf{P}}\big[G^{(n)}(X^n)\big] \geq \max_{\mathsf{Q}}\big[R(D,\mathsf{Q})-D(\mathsf{Q}\|\mathsf{P})\big].
\end{equation}
which provides us with a lower bound on $E(D)$.

As described in [19] 
the term $\log G^{(n)}(x^n)$ can be
thought of as the codeword length for a lossless entropy encoder that
operates on the adversary's strategy list $\mathcal{S}^{(n)}$ with
distortion constraint $D$. Thus, the combination of the entropy coding
and the corresponding strategy list creates a rate-distortion
code. Consequently, the average codeword length per input symbol
(i.e., $\frac{1}{n}\mathbb{E}_{\mathsf{Q}}\big[\log
G^{(n)}(X^n)\big]$) with respect to a source distribution $\mathsf{Q}$
cannot be smaller than the rate-distortion function $R(D,\mathsf{Q})$,
cf.~\eqref{eq:R_DQ}.

For a distribution $\mathsf{Q}$ defined on $\mathcal{X}$, we have
\begin{align}
\mathbb{E}_{\mathsf{P}}\big[&G^{(n)}(X^n)\big] = \sum_{x^n\in \mathcal{X}^n} \mathsf{P}^n(x^n)G^{(n)}(x^n) \notag\\
&=\sum_{x^n\in \mathcal{X}^n} \mathsf{Q}^n(x^n)\exp\Big(\log\frac{\mathsf{P}^n(x^n)G^{(n)}(x^n)}{\mathsf{Q}^n(x^n)}\Big)\notag\\
&=\mathbb{E}_{\mathsf{Q}}\Bigg[ \exp\Big(\log\frac{\mathsf{P}^n(x^n)G^{(n)}(x^n)}{\mathsf{Q}^n(x^n)}\Big)\Bigg]\notag\\
&\geqtop{a} \exp\Bigg(\mathbb{E}_{\mathsf{Q}}\Big[\log\frac{\mathsf{P}^n(x^n)G^{(n)}(x^n)}{\mathsf{Q}^n(x^n)}\Big]\Bigg)\notag\\
&=  \exp\Bigg(\sum_{x^n\in \mathcal{X}^n} \mathsf{Q}^n(x^n) \log\frac{\mathsf{P}^n(x^n)}{\mathsf{Q}^n(x^n)} \notag\\
&\qquad\qquad + \sum_{x^n\in \mathcal{X}^n} \mathsf{Q}^n(x^n)\log G^{(n)}(x^n)\Bigg)\notag\\
&\eqtop{b}  \exp\Bigg(-nD(\mathsf{Q}\|\mathsf{P}) 
+ \sum_{x^n\in \mathcal{X}^n} \mathsf{Q}^n(x^n)\log G^{(n)}(x^n)\Bigg), \label{eq:conv1}
\end{align}
where $(a)$ follows from Jensen's inequality, and $(b)$ is due to the fact that the random variable $X$ is i.i.d. and memoryless.

Next, define 
\begin{equation}
\lambda_j \coloneqq \sum_{\substack{x^n\in \mathcal{X}^n\\ \text{s.t. } G^{(n)}(x^n)=j}} \mathsf{Q}^n(x^n).
\end{equation}
Then, the second term of the exponential in \eqref{eq:conv1} writes as
\begin{equation}
\label{eq:conv2}
\sum_{x^n\in \mathcal{X}^n} \mathsf{Q}^n(x^n)\log G^{(n)}(x^n) = \sum_j \lambda_j\log j.
\end{equation}
Now, we are going to use a lossless entropy code for positive integers $j=\{1,2,\dots\}$ with probability distribution $\rho_j=\frac{C(\epsilon)}{j^{1+\epsilon}}$ for a fixed $\epsilon>0$. The constant $C(\epsilon)$ is chosen such that the probabilities satisfy $\sum_j \rho_j =1$. The average codeword length for this code is $\lceil \log_2\frac{1}{\rho_j}\rceil$ measured in bits [21], 
and as mentioned earlier, this entropy code  in combination with the return value of the counting function of the enrollment sequence $x^n$ (i.e., $j=G^{(n)}(x^n)$), resembles a rate-distortion code. Thus,
\begin{align}
R(D,\mathsf{Q}) &\leq \frac{1}{n\log_2 e} \sum_j\lambda_j\lceil\log_2\frac{1}{\rho_j}\rceil \notag\\
&\leq \frac{1}{n\log_2 e} \sum_j\lambda_j \left( 1+\log_2 \frac{j^{1+\epsilon}}{C(\epsilon)} \right) \notag\\
&= \frac{1}{n} \left(\log 2 + (1+\epsilon) \sum_j\lambda_j\log j - \log C(\epsilon)\right),
\end{align}
which leads to having
\begin{equation}
\label{eq:expRate}
\sum_j \lambda_j\log j \geq \frac{nR(D,\mathsf{Q}) - \log 2 + \log C(\epsilon)}{1+\epsilon}.
\end{equation}
Substituting \eqref{eq:expRate} and \eqref{eq:conv2} in \eqref{eq:conv1} and taking the logarithms from both sides, we get
\begin{align}
\frac{1}{n}\log\mathbb{E}_{\mathsf{P}}\big[G^{(n)}(X^n)\big] &\geq -D(\mathsf{Q}\|\mathsf{P}) + \frac{R(D,\mathsf{Q})}{1+\epsilon}\notag\\
&\;\;\quad - \frac{\log 2 - \log C(\epsilon)}{n(1+\epsilon)}.
\end{align}
Since the left-hand side is not dependent on the value of $\epsilon$ and the distribution $\mathsf{Q}$, taking the limit as $n\to\infty$ on both sides, then the limit as $\epsilon\to 0$ on the right-hand side, and finally the maximum over all possible distributions $\mathsf{Q}$ on the right-hand side, we obtain \eqref{eq:converse}. This completes the converse proof.\QEDA


\vspace{6pt}

\section{Authentication with Side Information at the Adversary} 
\label{sec:auth_SI}
Next, we consider an extension of the previous setup where we assume that the adversary may have access to some additional side information correlated with the enrollment. In particular, consider a pair of i.i.d.\ random sequences $X^n$ and $Z^n$ generated with joint distribution $\mathsf{P}$ defined on $\mathcal{X}\times \mathcal{Z}$. Here, $X^n$ represents the enrollment and $Z^n$ the side information available at the adversary.

The adversary takes advantage of this side information to impersonate the legitimate user by producing a guess of the enrollment sequence $x^n$ that falls within the distortion limit $D$. That is, for each measurement length $n$, the adversary has a strategy $\mathcal{S}^{(n)}_Z(z^n)\subseteq\mathcal{Y}^n$ which is an ordered set of $n$-length guesses $y^n_z(j)$ with $j=\{1,2,\dots\}$, i.e., 
\begin{equation}
\mathcal{S}^{(n)}_Z(z^n)\coloneqq\big\{y^n_z(1),y^n_z(2),\dots\big\}.
\end{equation}
With this strategy, the adversary produces a sequence of guesses until a guess $y^n_z(j)$ is found such that 
\begin{equation}
d^{(n)}\big(x^n,y^n_z(j)\big)\leq D,
\end{equation}
which leads to successful deception of the authentication system. 

The counting function $G^{(n)}(\cdot|z^n)$ of the adversary is now defined as the function that, given the side information $z^n\in \mathcal{Z}^n$, for each enrollment sequence $x^n\in\calX^n$ returns the index $j$ of the first guess $y^n_z(j)\in\mathcal{S}_Z(z^n)$ such that $d^{(n)}(x^n,y^n_z(j))\leq D$. If no such codeword exists, it returns $G^{(n)}(x^n|z^n)=\infty$.

\begin{definition}
The deception exponent with side information at the adversary under the distortion constraint $D$ is defined as
\begin{equation}
E_Z(D) \coloneqq \liminf_{n\to\infty}\frac{1}{n}\min_{\mathcal{S}^{(n)}_Z}\log\mathbb{E}_{\mathsf{P}}\big[G^{(n)}(X^n|Z^n)\big].
\end{equation}
\QEDD
\end{definition}

Similar to Section~\ref{sec:main_result}, we have the following characterization of the deception exponent with side information at the adversary,
which is again a special case of the results in [19] 
applied to the authentication setting.
\begin{theorem}
\label{th:main_res_SI}
The deception exponent with side information at the adversary under the distortion constraint $D$ is
\begin{equation}
E_Z(D) = \max_{\mathsf{Q}}\big[R_{X|Z}(D,\mathsf{Q})-D(\mathsf{Q}\|\mathsf{P})\big],
\end{equation}
where $\mathsf{Q}$ is a joint distribution defined on $\mathcal{X}\times \mathcal{Z}$. The term $R_{X|Z}(D,\mathsf{Q})$ is the rate-distortion function with side information available at both the encoder and the decoder [22] 
defined as 
\begin{equation}
\label{eq:R_DQ_SI}
R_{X|Z}(D,\mathsf{Q}) = \min_{\mathsf{P}(y_z|x,z):\mathbb{E}[d(X,Y_z)]\leq D} I(X;Y_z|Z).
\end{equation}
\QEDB
\end{theorem}

This result can be proved with similar steps taken in the proof of
Theorem~\ref{th:main_res}, paralleling the corresponding proof in [19],
using the rate-distortion function with side
information and the following extension of the type covering lemma,
cited from [19].
\begin{lemma}
Let $\mathcal{Q}^{(n)}(\mathcal{X},\mathcal{Y})$ denote the set of all joint types on $\mathcal{X}\times \mathcal{Y}$ of $n$-length sequences $(X^n,Z^n)\in\mathcal{X}^n\times \mathcal{Z}^n$. For any distortion measure on $\mathcal{X}\times \mathcal{Y}$, any joint type $\mathsf{Q}\in\mathcal{Q}^{(n)}(\mathcal{X},\mathcal{Y})$, and any distortion level $D\geq 0$, there exists a set $\mathcal{B}_Z(z^n)\subseteq\mathcal{Y}^n$ such that 
\begin{equation}
d^{(n)}(x^n,\mathcal{B}_Z(z^n))\coloneqq\min_{y^n_z\in\mathcal{B}_Z(z^n)} d^{(n)}(x^n,y^n_z)\leq D,
\end{equation}
for every $x^n\in T_V^n(z^n)$, and
\begin{equation}
\label{eq:typ_covering_cond}
\frac{1}{n}\log |\mathcal{B}_Z(z^n)|\leq R_{X|Z}(D,\mathsf{Q}) + \delta,
\end{equation}
where $\delta\to 0$ as $n\to\infty$. The notion $T_V^n(z^n)$ denotes the conditional type class defined as the set of all sequences $x^n\in \mathcal{X}^n$ having conditional type $V$ given $z^n\in \mathcal{Z}^n$, where $x^n$ and $y^n_z$ have a given joint type $\mathsf{Q}$. 
\QEDB
\end{lemma}


\section{Conclusions} 
\label{sec:conclusions}

In this paper, we have studied performance limits in authentication using information-theoretic arguments. The results provide valuable insights for the design and security assessment of an authentication system which is a key element in modern information-based systems. %
In particular, we considered authentication in a system where an adversary tries to launch an impersonation attack by guessing the enrollment sequence of a legitimate user stored in the database. %
This allowed us to apply the results from [19], 
which showed a relation between the authentication problem and the rate-distortion theory, and completely characterized the minimum number of required guesses (on average) by the adversary for a successful attack in terms of the deception exponent. 

We note that our model addresses a passive authentication system and an adversary with no knowledge of the legitimate user. While in practice, the system can take a more active role in the authentication process, e.g., by setting a limit on the number of guesses, we believe that these results can be used in the design of the authentication systems as they characterize the minimum required complexity 
of an adversary for successful deception in the system.

\section*{Acknowledgment} 
\label{sec:acknowledgement}
In this revised version we clarify the referencing made in the
conference version to the work in [19]. 
We apologize
to the authors of [19] 
as well as to the reviewers, that our
previous referencing was not very clearly pointing out the fact that
our work is a direct application of [19] 
to the authentication problem. We also acknowledge Neri Merhav for pointing
out the imprecise referencing in the conference version.



\bibliographystyle{IEEEtran}

\end{document}